\documentclass[sigconf, authorversion]{acmart}
\usepackage{lipsum}
\usepackage{booktabs}

\AtBeginDocument{%
  }

\copyrightyear{2025}
\acmYear{2025}
\setcopyright{rightsretained}
\acmConference[RecSys '25]{Proceedings of the Nineteenth ACM Conference on Recommender Systems}{September 22--26, 2025}{Prague, Czech Republic}
\acmBooktitle{Proceedings of the Nineteenth ACM Conference on Recommender Systems (RecSys '25), September 22--26, 2025, Prague, Czech Republic}
\acmDOI{10.1145/3705328.3748135}
\acmISBN{979-8-4007-1364-4/2025/09}

\begin{document}

\title[Suggest, complement, inspire: story of Two Tower recommendations at Allegro.com]{Suggest, complement, inspire: \\ story of Two Tower recommendations at Allegro.com}







\author{Aleksandra Osowska-Kurczab}
\authornote{All authors contributed equally to this research.}
\email{aleksandra.kurczab@allegro.com}
\affiliation{%
  \institution{Allegro.com}
  \country{Poland}
}
\orcid{0000-0001-5764-522X}

\author{Klaudia Nazarko}
\authornotemark[1]
\orcid{0009-0009-4133-211X}
\email{klaudia.nazarko@allegro.com}
\affiliation{%
  \institution{Allegro.com}
  \country{Poland}
}

\author{Mateusz Marzec}
\authornotemark[1]
\orcid{0009-0007-9859-1791}
\email{mateusz.marzec@allegro.com}
\affiliation{%
  \institution{Allegro.com}
  \country{Poland}
}

\author{Lidia Wojciechowska}
\orcid{0009-0009-5207-681X}
\affiliation{%
  \institution{Allegro.com}
  \country{Poland}
}

\author{Eliška Kremeňová}
\orcid{0009-0004-2333-7367}
\affiliation{%
  \institution{Allegro.com}
  \country{Czech Republic}
}

\renewcommand{\shortauthors}{Osowska-Kurczab et al.}

%
\begin{abstract}
Building large-scale e-commerce recommendation systems requires addressing three key technical challenges: (1)~designing a universal recommendation architecture across dozens of placements, (2)~decreasing excessive maintenance costs, and (3)~managing a highly dynamic product catalogue. This paper presents a unified content-based recommendation system deployed at Allegro.com, the largest e-commerce platform of European origin. The system is built on a prevalent Two Tower retrieval framework, representing products using textual and structured attributes, which enables efficient retrieval via Approximate Nearest Neighbour search. We demonstrate how the same model architecture can be adapted to serve three distinct recommendation tasks: similarity search, complementary product suggestions, and inspirational content discovery, by modifying only a handful of components in either the model or the serving logic. Extensive A/B testing over two years confirms significant gains in engagement and profit-based metrics across desktop and mobile app channels. Our results show that a flexible, scalable architecture can serve diverse user intents with minimal maintenance overhead.
\end{abstract}


  


\maketitle

\section{Introduction}
\label{sec:introduction}
Allegro is a major Central European e-commerce marketplace where over 20 million active buyers connect with more than 150 thousand sellers monthly to discover and purchase products. Recommendations are pivotal in driving organic and sponsored content discovery, with a significant share of attributed Gross Merchandise Value (GMV) and advertising revenue. Managing dozens of recommendation placements across the user journey introduces challenges in maintaining a universal and coherent system. Additionally, the platform's dynamic catalogue -- with its vast array of products, sellers, and users -- introduces complexities such as cold-start and long-tail distributions, highlighting the need for scalable and adaptable recommendation strategies.

A common approach to solving a recommendation problem is content-based filtering \cite{cbf}, which suggests items to users based on the intrinsic attributes of those items and historical user preferences. It matches item features to a user's profile, effectively framing recommendation as an information retrieval task focused on content similarity \cite{yt-dnn-recommender}. Alongside collaborative filtering \cite{ncf}, this method serves as a fundamental candidate generator in large-scale recommendation systems \cite{yt-dnn-recommender}.

The Two Tower (TT) model \cite{dlrm, item2item} is prominent in industrial recommendation systems due to its balance between predictive effectiveness and serving efficiency. It encodes inputs -- typically a \textit{query} representing user context and a \textit{target} representing an item from the catalogue -- into a shared embedding space using a deep learning model (DLRM), where relevance is inferred via the dot product of embeddings. Given the scale of large e-commerce platforms \cite{wang_personalized_2021}, exact similarity search becomes computationally infeasible; consequently, the combination of a DLRM encoder and Approximate Nearest Neighbour (ANN) indexing \cite{faiss, scann} is crucial for effective TT-based content recommendations.

With a wide range of placements to populate, making dedicated solutions for each scenario is impractical and expensive \cite{efficiency_effectiveness}. To address this, the industry usually turns to one of two paths: either large foundation models \cite{360Brew, HSTU}, which require sophisticated infrastructure for serving, or many domain-specific models \cite{Hao2020, Pal_2020}, which then result in high maintenance costs and complexity. We propose an architectural design to alleviate the issue:
\begin{itemize}
    \item we present a unified platform architecture tailored to content-based recommendations and demonstrate its effectiveness across three seemingly distinct tasks: similarity, complementary and inspirational recommendations (Fig.~\ref{fig:bike-reco}).
    \item we generalise these tasks by purposefully redefining the complementary and inspirational recommendation problems as variants of similarity search.
    \item we summarise insights from two years of continuous A/B testing on the Allegro platform to highlight the business impact of the proposed solution.
\end{itemize} 

\section{Methods}
\label{sec:methods}

\subsection{Similarity search with the Two Tower}
Two Tower model (Similarity-TT) is a canonical application of DLRM-based retrieval \cite{yt-dnn-recommender, item2item, sampling_bias_corrected} deployed at Allegro for similarity search tasks (e.g. retrieving substitute or similar products). It is trained to maximise the similarity between vector representations of query and target items from the product catalogue (item-to-item regime \cite{item2item}). Due to the massive amount of products (in a scale of hundreds of millions), the task is formulated as a classification problem with sampled softmax loss function \cite{yt-dnn-recommender} and mixed negative sampling strategy \cite{mixed-negative-sampling}. 

The high amount of volatile products makes training a distinct and learnable embedding for each product ID challenging and prone to overfitting \cite{one_epoch_phenomenon}. To address this, each product is represented by its content features, such as title, price, and category (with a hierarchical taxonomy). Each feature is transformed into a low-dimensional vector via a dedicated embedding table. All feature vectors are concatenated, passed through a multi-layer perceptron (FC), and L2-normalised. We apply weight tying between the query and target towers to enhance training speed and stability, creating a shared component collectively called the Product Encoder.

The architecture of the service has a distinct separation into offline and online components. Offline processing consists of data preparation, model training, evaluation, and index building. Training data comprises co-viewed product pairs, filtered to retain only pairs meeting a minimum co-occurrence threshold. Given the model's lightweight architecture, training can be completed efficiently on a machine with a single GPU (NVIDIA T4 16GB). Our models are retrained regularly, and ANN indexes are refreshed daily. The Faiss library \cite{faiss} handles index creation and online serving. Recommendations are served in real time with millisecond-level latency. The online service processes an incoming request by encoding the associated product using the Product Encoder and retrieving similar candidates from the indexed product catalogue.

Representing products solely based on their content features provides several advantages within Allegro's dynamic environment. The content-based recommendations complement existing collaborative filtering models and effectively mitigate the cold-start problem by generating representations for new products directly from their features. The lightweight content-based model eliminates the need for extensive product ID embedding tables, improving computational efficiency for offline training and online serving.

\subsection{Complementary Two Tower}
Complementary Two Tower (Complementary-TT) aims to provide supplementary product recommendations (e.g. tennis balls to tennis rackets). It is achieved by modifying the Product Encoder while keeping the same serving architecture as in similarity-TT.

\begin{figure}
    \centering
    \includegraphics[alt={Figure 2 displays a comparison of two proposed encoder architectures for product recommendation. On the left, labelled "Similarity-TT/Inspirational-TT," two identical encoder blocks are shown. Each block takes "Target embedding" and "Query embedding" as input, which are represented as stacks of colored rectangles indicating different feature types: blue for price, yellow for category, orange for seller, and red for title. These embeddings pass through a fully connected (FC) layer and L2 normalisation layer before being combined by an element-wise multiplication symbol enclosed in a circle. On the right, labelled "Complementary-TT," a similar encoder block processes "Target embedding," but the "Query embedding" is first passed through a "Query features" block that appears to concatenate the different feature embeddings. This combined query embedding then goes through an FC and L2 normalisation layers. Additionally, there is a "Complementary projection" block that takes the category feature of the query (yellow rectangle) and outputs a "complementary category" embedding (green rectangle). The outputs of the FC layer and the complementary projection are then combined via element-wise multiplication. A legend on the right explains the colour coding for each feature type and introduces green for "complementary category."},width=1\linewidth]{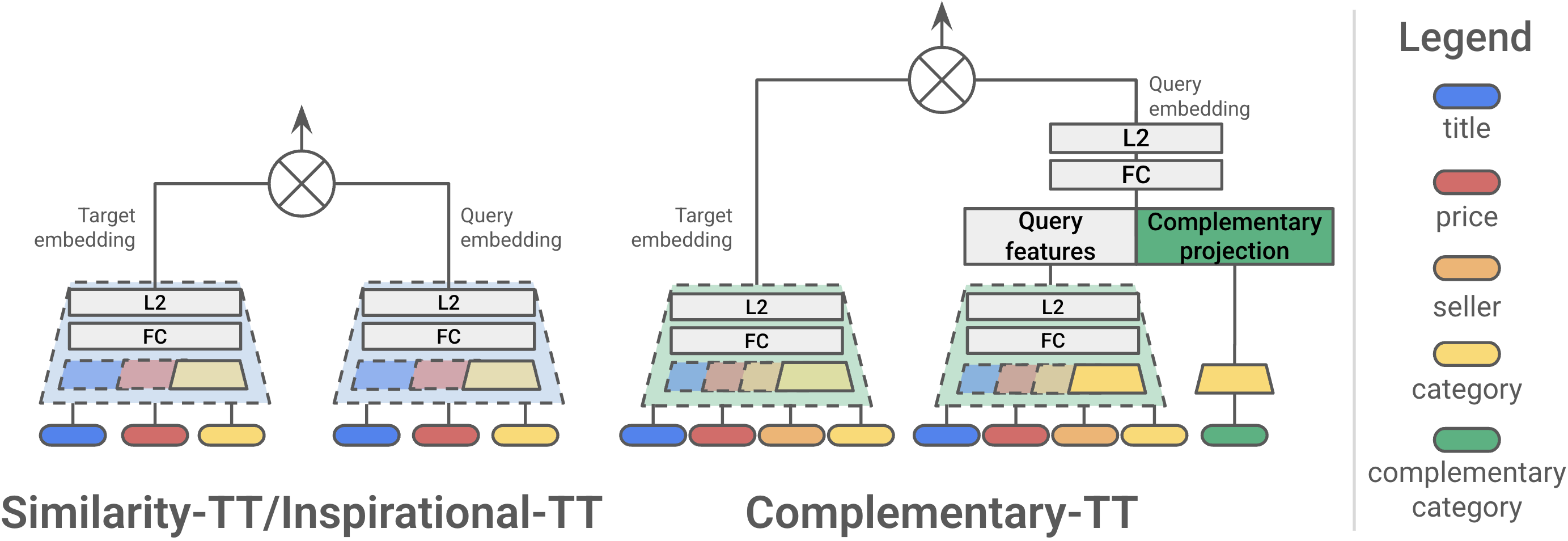}
    \caption{Comparison of two proposed TT architectures.}
    \label{fig:TT-architectures}
\end{figure}

The modifications are implemented in the query tower, with the target tower remaining unchanged (Fig.~\ref{fig:TT-architectures}).
The query tower takes an additional input, a complementary categories mapping (one-to-many), derived from the statistical models fit on co-purchase data, external annotations and domain knowledge.
Firstly, the target category is taken from the mapping and embedded with the same embedding tables as category inputs in Product Encoder. 
Then, the query product embedding is concatenated with the target category embedding and used as the final representation of the query \cite{Hao2020}. 
Inspired by \cite{Bibas_2023}, the loss function is enhanced with the target category reconstruction error to enforce the correctness of the target category embedding.
Besides the default Similarity-TT input features, the seller feature is used (due to co-purchase incentives, users tend to buy products from the same seller in a single transaction). 
The model is trained on co-purchase pairs, filtered to examples that follow a complementarity relation heuristic.

The online serving infrastructure remains the same as in the Similarity-TT model, with the only modification being the querying of the model with both the query product and the target complementary category. When the complementary categories mapping points to multiple target categories, the resulting groups of candidates are interleaved to enhance visual appearance of the carousel.

\begin{figure}
    \centering
    \includegraphics[alt={A pink children's bicycle is shown as the "Query product." To its right are three rows displaying product recommendations. The top row, labelled "Similar products," features five other children's bicycles in different colours: purple, blue, light blue, white with pink accents, and red. The middle row, labelled "Complementary products," displays a black helmet, black knee and elbow pads, a light blue helmet, a maroon helmet, and a black elbow pad. The bottom row, labelled "Inspirational products," includes a red and black light set, a pink and purple water bottle and cage, pink and purple handlebar streamers, a pink bicycle bell, a white wicker basket, and two small decorative bicycle windmills.},width=0.85\linewidth]{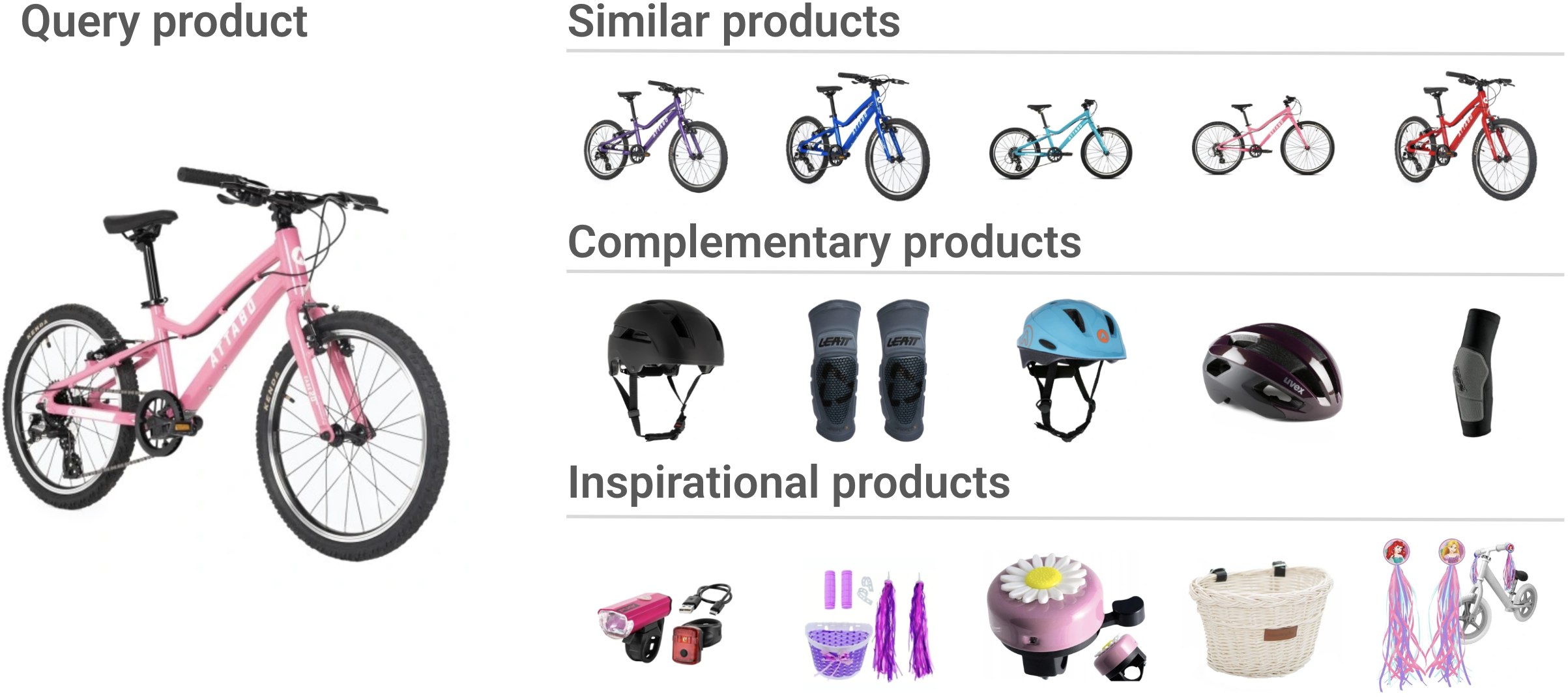}
    \caption{Recommendations generated by Similarity-TT, Complementary-TT and Inspirational-TT models.}
    \label{fig:bike-reco}
\end{figure}

\subsection{Inspirational Two Tower}
The goal of the Inspirational Two Tower (Inspirational-TT) algorithm is to encourage exploration by suggesting personalised products that are diverse and relevant to the user's browsing history. Such recommendations are obtained using the Similarity-TT Product Encoder with hierarchical ANN indexes to enable controllable diversification (similarly to \cite{Pal_2020}).

A hierarchical ANN index is constructed as follows: products are embedded with the Product Encoder, and then the k-means clustering algorithm is applied to product representations to generate $k$ distinct clusters (second-level indexes). Each cluster is represented by the centroid and added to the top-level index.
The user is represented by the last 100 product views from the past 7 days, aggregated into categories. The most recently viewed product from a given category serves as a category representative.
During the inference, the Product Encoder is queried with those representatives. Next, the top-level index is queried with the obtained representations, and $n$ closest clusters per category are selected: the more clusters, the more diverse the results. To avoid results that are too similar to the query, $l$ closest clusters may be skipped. In the third step, the most similar products to each category are fetched from second-level indexes and interleaved as the list of candidates.

While the Product Encoder architecture remains untouched, the online infrastructure must be adjusted by implementing hierarchical ANN indexes and introducing the aggregation of users' recently viewed products.

\section{Results}
\label{sec:results}
Recommendation systems at Allegro.com operate at production scale, powering online services with a throughput of 20k requests per second and 40ms p99 CPU latency. Models serve to address a variety of user intents, from exploration to purchase decisions. One recommendation system can be used to produce outputs for different recommendation scenarios (Fig.~\ref{fig:bike-reco}). With the query product being a bike, the regular TT model retrieves very similar products, mainly bicycles of the same brand and model, varying only in colour. The Complementary-TT model returns recommendations of supplementary products, such as helmets and knee pads. Results obtained from the Inspirational-TT model are characterised by more diversity: they are visually appealing and loosely connected accessories, such as bicycle bells, lamps, and decorations.

\subsection{Product related recommendations}

Product page remains one of the most prominent placements for recommendations at Allegro.com. It showcases products and enables navigation via recommendation carousels. Both Similarity-TT and Complementary-TT models were evaluated here as fallback candidate generators to collaborative-filtering (co-viewed and co-purchased model respectively). A/B tests were run with division to desktop and mobile app traffic.

Similarity-TT model was evaluated in the carousel "Others also viewed", which serves to present substitute products. Complementary-TT was tested in the carousel "Order in one parcel," which helps users find additional products from the same seller to complement their order and reach a minimal order value for free delivery. Due to business priorities, the latter test was run in the Sports, Travel and Fashion departments.
The two primary metrics for online evaluation were Click-Through Rate (CTR) and GMV per visit. The first metric evaluates user engagement, while the second measures financial profits delivered by the tested solution.

\begin{table}
    \centering
    \Description[Table 1 presents the A/B test results comparing the performance of the Similarity-TT and Complementary-TT models on a product page.]{The table shows the percentage change from a baseline metric for both Click-Through Rate (CTR) and Gross Merchandise Value (GMV), evaluated separately for mobile app and desktop users. An upward arrow next to CTR and GMV indicates that higher values are better. An asterisk (*) next to a percentage indicates statistical significance at the 0.01 level. For the mobile app, Similarity-TT shows a +2.11\% increase in CTR and a +0.13\% increase in GMV. Complementary-TT shows a +1.62\% increase in CTR (statistically significant) and a +0.09\% increase in GMV. For desktop, Similarity-TT shows a +2.37\% increase in CTR (statistically significant) and a +0.29\% increase in GMV. Complementary-TT shows a +1.06\% increase in CTR (statistically significant) and a +0.31\% increase in GMV.}
    \caption{A/B test results of Similarity-TT and Complementary-TT on product page. Values indicate \% change from the baseline, $\ast$ denotes statistical significance at the 0.01 level.}
    \begin{tabular}{ccccc}
         \toprule
         & \multicolumn{2}{c}{mobile app} & \multicolumn{2}{c}{desktop} \\
         \cmidrule{2-5}
         model  &CTR $\uparrow$  & GMV $\uparrow$  & CTR $\uparrow$  & GMV $\uparrow$ \\
         \midrule
         Similarity-TT  &+2.11\% $\ast$  &+0.13\%  &+2.37\% $\ast$  &+0.29\% \\
         Complementary-TT  &+1.62\% $\ast$  &+0.09\%  &+1.06\% $\ast$  &+0.31\% \\
         \bottomrule
    \end{tabular}
    \label{tab:ab-test-product-page}
\end{table}

Both TT and Complementary-TT positively influence CTR, suggesting that content-based models can capture hidden product relations and recommend items well suited to user needs (Table~\ref{tab:ab-test-product-page}). Increases in GMV per visit are mainly visible in desktop traffic, which can be explained by the differences in usage patterns between desktop and mobile devices.

\subsection{Inspirational recommendations}
Highly relevant product page recommendations perfectly fit customers with a purchase intent. However, the product page should also provide incentives to explore more and engage users looking for inspiration. Enhancing the product page with an Inspirational-TT recommender in the "How about..." section was the subject of an A/B test on desktop traffic. Two test variants differed in how inspirational recommendations were displayed: as an additional carousel vs an infinite feed. In both variants, models were queried with a single, currently viewed product (without the extended context of previously viewed products).

The focus of the A/B test was to improve engagement metrics, such as the Call To Action (CTA) and Conversion Rate (CVR), which calculate the ratio of users who engaged with the carousel. In addition, two auxiliary metrics were evaluated: exit rate and bounce rate, which measure the ratio of users who abandoned the page.

\begin{table}
    \centering
    \Description[Table 2 presents the A/B test results of introducing inspirational recommendations on the product page for desktop traffic.]{The table shows the percentage change from a baseline metric for views, Click-Through Rate (CTA), Conversion Rate (CVR), bounce rate, and exit rate. Upward arrows next to CTA and CVR indicate higher values are better, while downward arrows next to bounce rate and exit rate indicate lower values are better. An asterisk (*) next to a percentage indicates statistical significance at the 0.01 level, and the best result in each column is bolded. For the "carousel" view, the changes are +3.12\% for CTA, +1.38\% for CVR, -4.09\% for bounce rate, and -1.82\% for exit rate, all statistically significant. For the "infinite feed" view, the changes are +4.15\% for CTA, +2.22\% for CVR, -5.74\% for bounce rate, and -1.66\% for exit rate, which are statistically significant. The "infinite feed" shows the best results for CTA, CVR and bounce rate, while the "carousel" view shows the best result for exit rate.}
    \caption{A/B test results of introducing inspirational recommendations on product page (desktop traffic). Values indicate \% change from the baseline, $\ast$ denotes statistical significance at the 0.01 level, and the best result is bolded.}
    \begin{tabular}{ccccc}
         \toprule
         view & CTA $\uparrow$ & CVR $\uparrow$ & bounce rate $\downarrow$ & exit rate $\downarrow$ \\
          \midrule
         carousel  & +3.12\% $\ast$ &+1.38\% $\ast$ &-4.09\% $\ast$ & \textbf{-1.82}\% $\ast$ \\
         infinite feed  & \textbf{+4.15}\% $\ast$ & \textbf{+2.22}\% $\ast$ & \textbf{-5.74}\% $\ast$ &-1.66\% $\ast$ \\
         \bottomrule
    \end{tabular}
    \label{tab:ab-test-infinite-feed}
\end{table}

Presenting inspirational content on the product page led to substantial improvements in user engagement (Table~\ref{tab:ab-test-infinite-feed}). The CTA increased by 3.12\% in the carousel view and by 4.15\% in the infinite feed layout. It indicates that inspirational recommendations capture user attention and encourage them to explore more.

\section{Conclusions}
\label{sec:conclusions}
This work demonstrates that Two-Tower system architecture for similarity search can be easily adjusted for serving complementary or inspirational content. Extensive A/B tests show improvements in both engagement and profit-based metrics while maintaining minimal maintenance costs. This proves that the proposed solution can scale and adapt to numerous recommendation placements, fulfilling diverse user intents across their journey on the platform. Nonetheless, performance depends on content features quality, and deployment success requires coherence between encoder and indexing mechanism. Future work involves integrating user context into the model and evaluating its production impact.




\section*{Presenters' bio}
\textbf{Aleksandra Osowska-Kurczab, PhD} is a Machine Learning Manager leading the research team developing retrieval and ranking models deployed in e-commerce applications at Allegro.com. Her research interests include representation learning and robustness. \\
\textbf{Klaudia Nazarko} is a Machine Learning Research Engineer at Allegro.com, where she works on e-commerce recommender systems. Her work is focused on personalised retrieval and ranking. \\
\textbf{Mateusz Marzec} is a Machine Learning Research Engineer at Allegro.com, specialising in machine learning for e-commerce recommendations. Daily, he works on improving personalisation in retrieval systems.

\begin{acks}
This work is the result of a collaborative effort in the area of recommendation systems at Allegro.com for the past 2 years. We thank the alumni researchers: Michał Bień, Elwira Hołowko, Piotr Januszewski, Marcin Cylke for their foundational contributions to ML projects, and the engineering team: Maciej Arciuch, Mateusz Lamecki, Jakub Demianowski, Krzysztof Szczepański, Uladzislau Dziuba for their implementation support. 
\end{acks}

\bibliographystyle{ACM-Reference-Format}
\bibliography{bibliography}

\end{document}